\documentstyle[12pt]{article}
\title{{\Large {\bf 
Effective theory of  NN interactions in a separable 
representation\bigskip}}}
\author{B. Krippa and B.L. G. Bakker \\
 Department of Physics and Astronomy, Free University, 
Amsterdam,\\ De Boelelaan 1081, 1081 HV Amsterdam.\\}
\begin{document}
\maketitle
\vspace{1cm}
\begin{abstract}
We consider the effective field theory of the NN system
in a separable representation. The pionic part of 
the effective potential is included nonperturbatively
and approximated by a separable potential.
The use of a separable representation allows for the explicit 
solution of the Lippmann-Schwinger equation and a consistent
renormalization procedure. The phase shifts in the $^1S_0$ 
channel are calculated to subleading order.

\end{abstract}

\vskip0.9cm

\section{Introduction}
\label{intro}
During the last few years the effective field theory (EFT) has
extensively been used for the study of the NN interactions. The
activity in this field was inspired by the Weinberg proposal
\cite{We91} that the EFT approach could be useful in low energy nuclear
physics. Since Weinberg's original paper many aspects of this problem
have been discussed \cite{Ka96}.

Contrary to more phenomenological models of hadron interactions EFT
allows for systematic expansion of the scattering amplitude order by
order and the possibility to estimate a priory the  anticipated errors
at each order of the expansion using power counting rules.  However,
while applied to the two-nucleon systems, EFT encounters a serious
difficulty which is due to the existence of the extremely large the
S-wave scattering length (compared to the pion Compton wavelength).
Thus, it turns out that the EFT description of the NN forces must be
nonperturbative to incorporate this large scale.  In the original work
\cite{We91} Weinberg proposed to apply counting rules to the
irreducible   diagrams  in order to construct the effective potential
which is to be iterated in the Lippmann-Schwinger (LS) equation.
Immediately one can see a complication . The corresponding effective
potential is highly singular. The origin of this singularity is the
local nature of the nucleon-nucleon coupling. In order to obtain
finite physical observables one needs to carry out  the procedure of
regularization and renormalization.  The issue of renormalization is
much more involved in the case of the nucleon-nucleon interaction as
compared to the standard perturbative situation where the
renormalization can be carried out for the set of individual Feynman
diagrams, using the standard textbook methods. For the problem in hand,
nonperturbative renormalization is required so that at every order the
divergences of the whole nonperturbative amplitude must be subtracted.

A  somewhat different way to construct the EFT of the NN-forces has
been proposed some time ago by Kaplan et al. \cite{Ka98}. The idea was
to sum up a certain subclass of the leading order diagrams, given by
the lowest order contact interactions.  The rest, including the higher
order contact interactions and graphs with pions, can then be treated
perturbatively.  This approach is systematic, chirally symmetric, and is
formulated in such a way that chiral counting rules can be applied
directly to the nucleon-nucleon scattering amplitude. The leading
non-perturbative amplitude can be calculated in an analytic form,
allowing for the renormalization to be carried out in explicit and
transparent way.  The renormalization of the perturbative corrections
can be performed using the standard  methods of dealing with
divergencies of Feynman diagrams. However, the perturbative ``pionic
part'' of this approach seems to show rather slow convergence in some
particular channels \cite{Fl}, making the practical use of this
approach somewhat problematic.

In the Weinberg approach pion effects are treated to all order.  At
very low energies, when pion degrees of freedom can safely be
integrated out, the scattering amplitude can be derived analytically and
so no problem with the renormalization arises. In the more general case
of a potential consisting of the contact terms and a long-range one
pion exchange (OPEP) contribution, the analytic solution of the
three-dimensional LS equation is no longer possible and the problem
must be treated numerically. However, it is not at all clear how to
carry out the renormalization in such a situation. One notes that it is
not enough to regularize the integral part of the LS equation by
imposing a simple cut-off or using form-factors. In this case one is
still left with the bare couplings and the physical amplitude may
strongly depend on the value of the cut-off parameter. It contradicts
the renormalization group requirements, according to which \cite{Bi}
the physical NN-amplitude must be cut-off independent (at least up to
the order one is dealing with).  To remove the unwanted cut-off
dependence one needs to switch to renormalized effective couplings.
However, this is difficult to implement in the situation where the
analytical solution is not known.

In this paper we propose an approximate method of how to carry out the
renormalization if an exact  solution of the LS equation is 
not possible. Namely, we propose to use the approximate analytical solution,
which can be obtained if we represent the pionic part of the effective
Lagrangian by a sum of separable terms. In this case the integral
equation can be transformed into a matrix equation and an analytical
solution becomes possible. Then the renormalization can be carried out by
subtracting the loop integrals at some fixed kinematical point
$p^2 = -\mu^2$ and by replacing the bare constants with the running  
ones, depending on the point of subtraction. 

\section{Model}
\label{model}
We start from the standard nonrelativistic effective Lagrangian   
\begin{equation}
{\cal L}=N^\dagger i \partial_t N - N^\dagger \frac{\nabla^2}{2 M} N
- \frac{1}{2} C (N^\dagger N)^2\\ 
-\frac{1}{2} C_2 (N^\dagger \nabla^2 N) (N^\dagger N) +{\cal L_\pi}+ h.c. + \ldots.
 \label{eq.001}
\end{equation} 
Here ${\cal L_\pi}$ is the ``pionic'' part of the effective chiral
Lagrangian.  This Lagrangian leads to the following effective potential
for the $^1S_0$ NN scattering \cite{We91}
\begin{equation} V(\vec{p},\vec{p}') = C' + C_2 (\vec{p}^2
+ \vec{p}'^2) + V_{\pi}(\vec{p},\vec{p}'),
 \label{eq.002} 
\end{equation} 
where 
\begin{equation}
 C' =  C + \frac{g_{A}^2}{2f_{\pi}^2};\hskip1cm
  V_{\pi}(\vec{p},\vec{p}') = - \frac{\alpha_{\pi}}{\vec{q}^2 +
  m_{\pi}^2};\hskip1cm \alpha_{\pi} = \frac{g_{A}^2 m_{\pi}^2}{2 f_{\pi}^2},
 \label{eq.003} \end{equation} 
$\vec{q} = \vec{p} - \vec{p}'$, $g_{A} = 1.25$ and $f_{\pi}=132$ MeV
are the axial and pion decay constant respectively.  As mentioned above
the consistent numerical realization of the renormalization program in
the nonperturbative situation is a very difficult task (see for example
Ref.\cite{Ph}) therefore we adopt the strategy of an approximate
analytic solution of the LS equation allowing for the explicit
realization of the renormalization procedure. To achieve this goal we
represent the OPEP contribution by a sum of the separable terms. As we shall
henceforth limit our discussion to $S$-waves only, the matrix elements are
functions of the magnitudes of the momenta only. We write

\begin{equation}
 V_{\pi}(p,p')=\sum_{j=1}^{n}\alpha_{j} \eta_{j}(p)\eta_{j}(p') .
\end{equation}
One notes that, in principle,  $V_{\pi}(p,p')$ can be
parametrized with arbitrary accuracy but in this short letter we rather
would like to emphasize the issues related to renormalization in the
effective description of the NN interaction. So in practice we retain
only one term in a separable expansion. It turned out to be enough to
illustrate the main features of our approach. Of course, this is a
quite crude description of the OPEP, which approximate the exact pionic
part of the effective Lagrangian with an average error about 10-12$\%$
in the momentum region $0.4$ fm$^{-1} < p < 1.4$ fm$^{-1}$. We postpone the
detailed analysis of the  NN observables in the different partial waves
and spin-isospin channels until future publication.

After the separable approximation is substituted, the effective
potential can be represented in the following matrix form
\begin{equation}
 V^{\rm eff}(p,p')= \sum_{ij} g_i(p)M_{ij}(p)g_j(p') ,
\end{equation}
where
\begin{equation}
 g_i(p) = \left( \begin{array}{c} 1
 \\ p^2 \\ \eta_{1}(p) \end{array} \right)\quad
 {\rm and} \quad M_{ij}
 = \left( \begin{array}{clc} C' & C_2 & 0 \\ C_2 & 0 & 0 \\ 0 & 0 & \alpha_1 
\end{array} \right).
 \label{eq.004}
\end{equation}

The solution of the LS equation can be represented as 
\begin{equation}
 T(p,p';E)=g_i(p)\tau_{ij}(E)g_j(p') .
 \label{eq.005}
\end{equation}

We denote $\tau$ the $3 \times 3$ matrix containing the loop integrals
$I_{ij}(E)$, given by
\begin{equation}
 \tau(E)= \left[ 1 - M\, I(E)\right]^{-1}M ,
 \label{eq.006}
\end{equation}
where 
\begin{equation}
 I_{ij}(E) = \int^\infty_0
 \frac{d q q^2 }{2\pi^2} \frac{g_i(q) g_j(q)}{E + i \epsilon - E(q)}.
 \label{eq.007}
\end{equation}
The matrix $\tau(E)$ contains convergent and divergent integrals so
regularization and renormalization must be carried out. We use the
subtraction scheme smilar tho the one suggested suggested in \cite{Ge}.
Namely, all loop integrals are subtracted at some kinematical point
$p^2 = -\mu^2$. The  renormalized T-matrix is
\begin{equation}
 T^{\rm Reg}(p,p';E) =
 g_i(p)\; \tau^{\rm Reg}_{ij}(E)\; g_j(p') .
 \label{eq.008}
\end{equation}

In the following we will omit the superscript ``Reg'' implying that we
always work with the renormalized amplitude. After renormalization the
Low-energy effective constants become dependent on the renormalization
point $\mu$. The ``$\mu$ independence'' of the scattering amplitude is
provided by the renormalization group (RG) equations. Requiring that
$dT/d\mu$ = 0 and using the analytic expression for the T-matrix one
obtains the following set of RG equations for the leading order
coefficient $C$.

\begin{equation}
\frac{\partial C(\mu)}{\partial \mu} = \frac{C'^2 M}{4\pi}
 - 2\alpha \eta^2_1 (p) C' \frac{M}{4\pi} .
 \label{eq.009}
\end{equation}  

Neglecting the term with the form-factors $g_i(p)$ we arrive at the
variant of the RG equations first derived by Kaplan et al.~\cite{Ka98}
where pions were included perturbatively.  In the region where the
second term becomes nonnegligible, the pionic effect must be treated
in a nonperturbative manner.

\section{Numerical Results}
\label{results}

We used the exponential form of the separable form-factors to parametrize
the one-pion exchange potential
\begin{equation}
 \eta_1(p)= \exp(-\beta p) .
 \label{eq.010}
\end{equation}
The cut-off parameter $\beta$ and strengh parameter $\alpha$ are taken
to be $0.78$ fm and $1.73$ fm$^2$ respectively. 
The values of the effective constants used to calculate
the phase shifts are $C(m_\pi) = -3.2$ fm$^2$ and $C_{2}(m_\pi) = 2.5$
fm$^4$.  These values are to be compared with the chiral counting
rules, according to which $C_{2n} (\mu) \sim 4 \pi/(M \Lambda^n
\mu^{n+1})$, where $\Lambda$ is the scale where chiral perturbation
theory breaks down.  Assuming $\Lambda \sim 300 - 400$ MeV, one finds
that the values of the effective constants are indeed consistent with
the counting rules, although somewhat lower than thode obtained in
Ref.~\cite{Ka98}. One notes that it is \ hard to compare the effective
constants obtained in different regularization schemes, since the
coupling is known to be a scheme dependent quantity.

The nonperturbative corrections due to the separable potential with the
form-factor $\eta_1(p)$ become noticeable at $p \sim 100$ MeV$/c$. It
agrees with the estimates obtained in \cite{CoH}.  Of course, the
precise region where pions become nonperturbative may somehow depend on
the concrete form-factors used, but the general tendency of the pion
effects to become too strong to be treated perturbatively at $p > 0.5$
fm$^{-1}$ seems to be quite robust, making the whole problem much more
complicated.

As already mentioned in this paper we focus on the $^1S_0$ channel and
calculate observables up to next-to-leading order. The main goal was to
develop a reasonable calculational scheme with consistent
renormalization procedure so we retained only one term in  the
separable expansion of OPEP contribution. Of course, it gives only
a crude parametrization of the long-range part of the effective
Lagrangian so that our comparison of the theoretical results with the
experimental phase shifts has somewhat illustrative character to
demonstrate the feasibility of the method proposed.  The results
obtained are shown in Table~\ref{tab.001}. 

\begin{table}
\begin{center}
\caption{Numerical results for the phase shifts $\delta(^1S_0)$ in degrees
obtained within the separable approximation to 
the OPEP compared to the results from the Nijmegen phase-shift analysis.
$T_{\rm Lab}$ is given in MeV and $p$ in fm$^{-1}$.}

\vspace{5mm}

\begin{tabular}{|r|r|r|r|}
\hline
 $T_{\rm Lab}$ & $p$ & $\delta_{\rm sep \; pot}$ & $\delta_{\rm Nijm}$ \\
\hline
 20 &  0.48 & 46.9 & 53.6 \\
 50 &  0.77 & 43.6 & 40.1 \\
 70 &  0.91 & 40.1 & 34.3 \\
 90 &  1.02 & 34.7 & 29.1 \\
 110 &  1.13 & 26,8 & 24.6 \\
 130 &  1.24 & 17.6 & 20.6 \\
 150 &  1.32 & 10.0 & 16.9 \\
 170 &  1.41 &  2.6 & 13.6 \\
\hline
\end{tabular}
\label{tab.001}
\end{center}
\end{table}

The deviation from the Nijmegen phase shifts \cite{Nijm} is about
12-15$\%$ on average in the kinematical  region $0.4$ fm$^{-1} < p <
1.35$ fm$^{-1}$. At lower momenta the pionic effects can either be
integrated out or safely treated perturbatively. At larger momenta the
next-next-to-leading order corrections like two-pion-exchange or
$O(p^4)$ contact terms become more and more important and must be
taken into account. The errors of the theoretical analysis are
comparable with those introduced by the separable
representation of the effective potential, so no significant additional
uncertainties are introduced by the loop integration. Therefore one
could hope that taking into account a few more  terms in the separable
expansion of the effective potential will bring the theoretical results
into better agreement with the experimental phase shifts. Work in this
direction is in progress.

In summary, we analyzed the problem of renormalization in the effective 
theory of the NN interaction when the perturbative chiral expansion is 
not valid. In Weinberg's approach, where pions are treated nonperturbatively,
the scattering amplitude can be found only numerically, making the procedure
of consistent renormalization difficult to implement. On the other hand, in the
approach proposed by Kaplan et al., pions are treated perturbatively, so that
renormalization can be carried out in the standard way. The latter approach 
however, shows rather slow convergence in some channels. The procedure
we propose is based on the approximate but nonperturbative treatment of
pionic effects based on a separable expansion of the long-range part of
the effective potential ans allowing for the renormalization to be carried out
in an analytic form. Our method gives a reasonable description of the
$^1S_0$ NN phase shifts in the laboratory-energy region up to $T_{\rm Lab} 
\sim 140$ MeV.

\end{document}